\documentclass[preprintnumbers,amsmath,amssymb,showpacs]{revtex4}
\usepackage{graphicx}% Include figure files
\usepackage{dcolumn}% Align table columns on decimal point
\usepackage{bm}% bold math
\usepackage{booktabs}
\usepackage{CJK}

\begin{document}

\title{Entanglement measure and quantum violation of Bell-type inequality for a family of four-qubit entangled states}

\author{Dong Ding$^{1,2}$, Yingqiu He$^{2}$, Fengli Yan$^2$}
 \email{flyan@hebtu.edu.cn}
\author{Ting Gao$^3$ }
 \email{gaoting@hebtu.edu.cn}
\affiliation {$^1$Department of Basic Curriculum, North China Institute of Science and Technology, Beijing 101601, China \\
$^2$ College of Physics Science and Information Engineering, Hebei Normal University, Shijiazhuang 050024, China \\
$^3$College of Mathematics and Information Science, Hebei Normal University, Shijiazhuang 050024, China}
\date{\today}

\begin{abstract}

By calculating entanglement measures and quantum violation of Bell-type inequality,
we reveal the relationship between entanglement measure and the amount of quantum violation for a family of four-qubit entangled states.
It has been demonstrated that the Bell-type inequality is completely violated by  these four-qubit entangled states. The plot of entanglement measure as a function of the expectation value of Bell operator  shows that entanglement measure first decreases and then increases smoothly with increasing quantum violation.

\end{abstract}

\maketitle

\section{Introduction}

As a tangible physical resource, quantum entanglement \cite{HHHH2009,GT2009} is one of the chief differences between quantum and classical mechanics. Without a doubt, the study of quantum entanglement is  a significant problem for the development of quantum information processing \cite{NC2000,Peres1996PRL77-1413,DVC2000,YGC2011,GH2011EPJD,HGY2012PRA86-062323,W2013PRA87-062316,W2014PRA90-033830,GYE2014}.
Entanglement measures (or entanglement monotones) are utilized for quantifying the amount of entanglement in a given state.
As is well-known, the concurrence \cite{concurrence1997, concurrence1998} is a popular measure for bipartite entanglements and the three-tangle \cite{three-tangle2000} is used to quantify tripartite quantum correlations. Furthermore,  many axiomatic entanglement measures are extended to the multipartite systems \cite{WC2001PRA63-044301,EB2001PRA64-022306,M2003PRA67-012108,OS2005PRA72012337,SS2013PRA87-022335}.

Bell inequality \cite{Bell1964}, as the oldest tool to detect entanglement, was designed to rule out local hidden variable models.
Motivated by the Bell's ground-breaking discovery,  recent years Bell-type inequalities have been generalized from two-particle to $n$-particle cases. Specially,  Clauser-Horne-Shimony-Holt inequality \cite{CHSH1969}, Mermin-Ardehali-Belinski\u{\i}-Klyshko inequality  \cite{Mermin1990,Ardehali1992,BK1993}, Werner-Wolf-\.{Z}ukowski-Brukner inequality \cite{WW2001,ZB2002}  have been obtained. Although there exist detection efficiency loophole and locality loophole, the generalized Bell inequalities \cite{WYKO2007PRA75-032332,LF2012, WZCG2013,DHYG2015CPB,HDYG2015EPL} are an important tool for the investigation of possible connections between entanglement and quantum nonlocality for multiparticle systems.
For example, based on the Svetlichny operator  \cite{Svetlichny1987}, Ghose \emph{et al} \cite{GSDRS2009} investigated the relationship between tripartite entanglement and nonlocality for the Greenberger-Horne-Zeilinger (GHZ) class, and then they generalized the result to four-qubit case \cite{GDSKS2010}.

We here focus on four-qubit system and consider a family of the entangled states
\begin{equation}\label{}
  \left|\psi(\theta)\right\rangle=
     \frac{\cos{\theta}}{2}(\left| {0000} \right\rangle - \left| {0101} \right\rangle  + \left| {1010} \right\rangle + \left| {1111} \right\rangle)
         + \frac{\sin{\theta}}{2}(\left| {0110} \right\rangle  + \left| {1001} \right\rangle - \left| {0011} \right\rangle  + \left| {1100} \right\rangle)
\end{equation}
with $0\leq\theta\leq\pi/2$, which belong to the states \cite{Yeo2006PRA74-052305}
\begin{equation}\label{}
  \left|\psi(\theta_1,\theta_2)\right\rangle=
     \frac{\cos{\theta_1}}{2}(\left| {0000} \right\rangle  + \left| {1111} \right\rangle)
         - \frac{\sin{\theta_1}}{2}(\left| {0011} \right\rangle  - \left| {1100} \right\rangle)
         - \frac{\cos{\theta_2}}{2}(\left| {0101} \right\rangle  - \left| {1010} \right\rangle)
         + \frac{\sin{\theta_2}}{2}(\left| {0110} \right\rangle  + \left| {1001} \right\rangle),
\end{equation}
with $0\leq\theta_1, \theta_2\leq\pi/2$.
More specifically, for $\theta_1=\theta_2$ we set $\left|\psi(\theta_1,\theta_2)\right\rangle=\left|\psi(\theta)\right\rangle$, and then for $\theta_1=\theta_2=\pi/4$, $\left|\psi(\pi/4)\right\rangle \equiv |\chi\rangle$, where $|\chi\rangle$ is a suitable candidate for four-qubit maximally entangled state \cite{YC2006PRL96-060502}.
We next calculate two kinds of four-qubit entanglement measures and discuss the violation of one of the four-qubit Bell-type inequalities, the Wu-Yeo-Kwek-Oh (WYKO) inequality \cite{WYKO2007PRA75-032332}, for the entangled states $\left|\psi(\theta_1,\theta_2)\right\rangle$.
Noting that the WYKO inequality is optimally violated by $|\chi\rangle$ but not violated by the GHZ state, here we further reveal that there exist a family of four-qubit entangled states $\left|\psi(\theta)\right\rangle$ for which the WYKO inequality is completely violated.
At last, we show a relationship between entanglement measure and quantum violation for this family of four-qubit entangled states.

\section{Entanglement measures for the states $\left|\psi(\theta_1,\theta_2)\right\rangle$}

Now, we first consider the four-qubit entanglement measures for the entangled states $\left|\psi(\theta_1,\theta_2)\right\rangle$.
Generally, for an $n$-qubit system, Wong and Christensen \cite{WC2001PRA63-044301} proposed a compact measure of pure states for even $n$, that is,
\begin{equation}\label{tau-n}
\tau_n(\psi)=|\langle\psi|\widetilde{\psi}\rangle|^{2},
\end{equation}
where $|\widetilde{\psi}\rangle=\sigma^{\otimes n}_{y}|\psi^{*}\rangle$, $\sigma_{y}$ is a Pauli matrix.
For the states $\left|\psi(\theta_1,\theta_2)\right\rangle$, it is easy to obtain
\begin{equation}\label{tau-4}
\tau_4[\psi(\theta_1,\theta_2)]=\sin^2(\theta_1-\theta_2)\sin^2(\theta_1+\theta_2).
\end{equation}

Also, more recently, we note that Sharma \emph{et al} \cite{SS2013PRA87-022335} defined an entanglement monotone to quantify four-qubit correlations as
\begin{equation}\label{}
\tau_{(4,8)} = 4|\sqrt{12I_{(4,8)}}|,
\end{equation}
where $I_{(4,8)}$ is a four-qubit invariant of degree $8$ expressed in terms of three-qubit invariants.
For the states $\left|\psi(\theta_1,\theta_2)\right\rangle$, we calculate $\tau_{(4,8)}$ and find
\begin{eqnarray}\label{tau-4-8}
\tau_{(4,8)}[\psi(\theta_1,\theta_2)] & =& 8\sqrt{3}|\{ 3[\frac{1}{6}(a_{0000}a_{1111}+a_{0011}a_{1100} - a_{0101}a_{1010} - a_{0110}a_{1001})^2  \nonumber \\
          & &- \frac{2}{3}(a_{0000}a_{1100}a_{0011}a_{1111} + a_{0110}a_{1010}a_{0101}a_{1001}) ]^2  \nonumber \\
          & &+ 16 a_{0000}a_{1100}a_{0110}a_{1010}a_{0011}a_{1111}a_{0101}a_{1001} \}^{1/2}|\nonumber\\
          &=& \frac{1}{2}\sqrt{[1+\cos(2\theta_1)\cos(2\theta_2)]^2+3\sin^2(2\theta_1)\sin^2(2\theta_2)},
\end{eqnarray}
where $a_{i_1i_2i_3i_4}$, $i_1,i_2,i_3,i_4 \in \{0,1\}$, are respectively nonzero terms with the expansion of the states $\psi(\theta_1,\theta_2)$ in computational  basis.

\begin{figure}
  \includegraphics[width=4in]{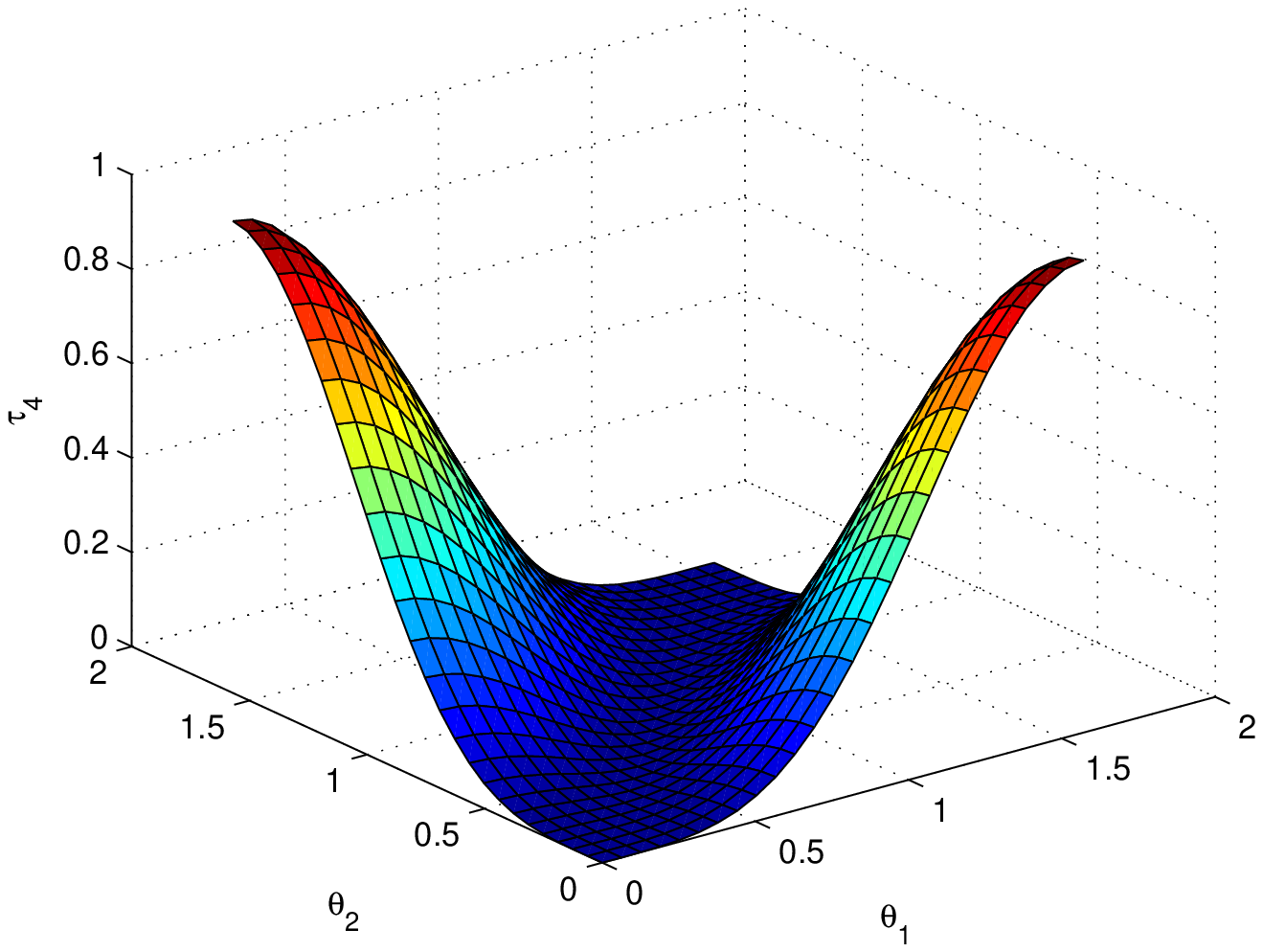}\\
  \caption{Plot of the entanglement measure $\tau_4$ as a function of $\theta_1, \theta_2$ for the states $\left|\psi(\theta_1,\theta_2)\right\rangle$.}
  \label{tau-4-theta-1-2}
\end{figure}

\begin{figure}
  \includegraphics[width=4in]{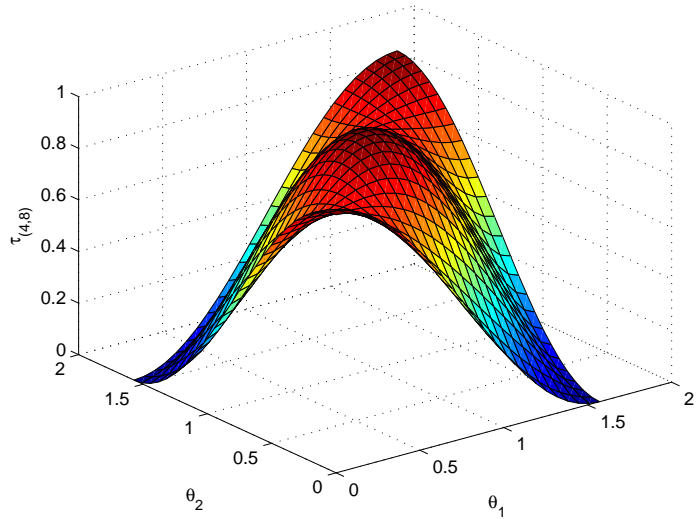}\\
  \caption{Plot of the entanglement measure $\tau_{(4,8)}$ as a function of $\theta_1, \theta_2$ for the states $\left|\psi(\theta_1,\theta_2)\right\rangle$.}
  \label{tau-4-8-theta-1-2}
\end{figure}

In order to gain an intuitive understanding of the two expressions, we plot the entanglement measures as a function of $\theta_1, \theta_2$ for the states $\left|\psi(\theta_1,\theta_2)\right\rangle$ respectively, as shown in Fig.\ref{tau-4-theta-1-2} and Fig.\ref{tau-4-8-theta-1-2}.
It is easy to see that for the states $\left|\psi(\theta_1,\theta_2)\right\rangle$ the entanglement measures (\ref{tau-4}) and (\ref{tau-4-8}) are very different.
Especially, for $\theta_1=\theta_2=\theta$, $0 < \theta < \pi/2$, we have
\begin{equation}\label{}
\tau_4[\psi(\theta)]=0,
\end{equation}
and
\begin{equation}\label{tau-4-8-theta}
\tau_{(4,8)}[\psi(\theta)] = \sqrt{\cos^4(2\theta)+\sin^2(2\theta)}.
\end{equation}
Thus only $\tau_{(4,8)}[\psi(\theta)]$ is allowed to vary with $\theta$.

\section{Entanglement measure and quantum violation for the states $\left|\psi(\theta)\right\rangle$}

Consider a Bell-type inequality for four-qubit systems, the WYKO inequality \cite{WYKO2007PRA75-032332}
\begin{equation}\label{BI}
|\langle B \rangle| \leq 2,
\end{equation}
where $\langle B \rangle$ represents the expectation value of Bell operator $B$ in some states with $B=A_1B_1C_1D_1+B_1C_2D_2+B_2C_1D_2-A_1B_2C_2D_1$,
$X_1$ and $X_2$ are respectively measurement operators on the locations of qubit $X$ with $X\in \{A,B,C,D\}$.
If one takes an appropriate set of experimental settings \cite{WYKO2007PRA75-032332} as $A_1=\sigma_x, B_1=\sigma_z, C_1=\sigma_z, D_1=\sigma_x, B_2=\sigma_y, C_2=\sigma_y$, and $D_2=\sigma_y$, then the Bell operator $B$  becomes 
\begin{equation}\label{}
B=\sigma_x\sigma_z\sigma_z\sigma_x+\sigma_0\sigma_z\sigma_y\sigma_y+\sigma_0\sigma_y\sigma_z\sigma_y-\sigma_x\sigma_y\sigma_y\sigma_x,
\end{equation}
where $\sigma_0$ is identity operator and $\sigma_i$ ($i=x,y,z$) is Pauli operator.

With a straightforward calculation, one can obtain the expectation value of $B$ in the states $\left|\psi(\theta_1,\theta_2)\right\rangle$, that is,
\begin{equation}\label{}
\langle\psi(\theta_1,\theta_2)|B|\psi(\theta_1,\theta_2)\rangle=[1+\cos(\theta_1-\theta_2)][1+\sin(\theta_1+\theta_2)].
\end{equation}
In particular, for $\theta_1=\theta_2=\theta$, we have
\begin{equation}\label{B-theta}
\langle\psi(\theta)|B|\psi(\theta)\rangle=2[1+\sin(2\theta)].
\end{equation}
Obviously, the inequality is violated by all the states $\left|\psi(\theta)\right\rangle$. Furthermore, for $\theta=\pi/4$, the inequality is maximally violated  with $\langle\chi|B|\chi\rangle=4$.

\begin{figure}
  \includegraphics[width=4in]{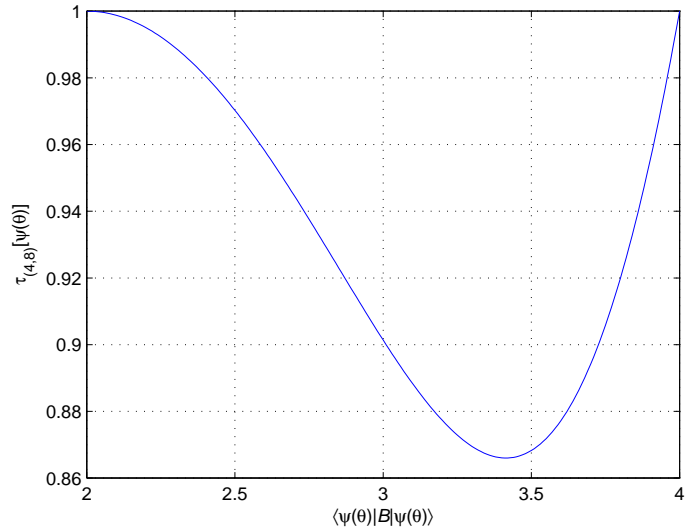}\\
  \caption{Plot of the relationship between $\tau_{(4,8)}[\psi(\theta)]$ and $\langle\psi(\theta)|B|\psi(\theta)\rangle$.}
  \label{tau-4-8-B}
\end{figure}

Based on our calculations for the states $\left|\psi(\theta)\right\rangle$, we  obtain a useful relationship between $\tau_{(4,8)}[\psi(\theta)]$ and $\langle\psi(\theta)|B|\psi(\theta)\rangle$. It reads
\begin{equation}\label{tau-theta}
\tau_{(4,8)}[\psi(\theta)] = \sqrt{1+(1-\langle\psi(\theta)|B|\psi(\theta)\rangle/2)^4-(1-\langle\psi(\theta)|B|\psi(\theta)\rangle/2)^2},
\end{equation}
as shown in Fig.\ref{tau-4-8-B}.
In comparison with the previous expressions \cite{GSDRS2009,GDSKS2010}, our results show that the entanglement measure varies smoothly with the value of quantum violation, and as the value of quantum violation increases, $\tau_{(4,8)}[\psi(\theta)]$ first decreases and then increases, in the argument interval $2 < \langle\psi(\theta)|B|\psi(\theta)\rangle \leq 4$. Obviously, the entanglement measure decreases to the minimum value $\tau_{(4,8)}[\psi(\pi/8)]=\sqrt{3}/2$ as $\langle\psi(\theta)|B|\psi(\theta)\rangle = 2+\sqrt{2}$, i.e. $\theta=\pi/8$.

\section{discussion and summary}

In summary, we have calculated two kinds of four-qubit entanglement measures for the states $\left|\psi(\theta_1,\theta_2)\right\rangle$ and shown that $\tau_{(4,8)}$ is a more suitable measure than $\tau_{4}$ for the present states. Furthermore, the entanglement measure and the quantum violation of Bell-type inequality for a family of four-qubit entangled states $\left|\psi(\theta)\right\rangle$ have been investigated. We revealed the relationship between entanglement measure and the amount of quantum violation of the WYKO inequality for the states $\left|\psi(\theta)\right\rangle$ in terms of the numerical calculation.
Especially for the state $|\chi\rangle$  corresponding to $\theta=\pi/4$, there exists an interesting result, i.e.,  the maximum entanglement measure $\tau_{(4,8)} = 1$ corresponding to the maximum violation 4.
Our results differ from the previous studies \cite{GSDRS2009,GDSKS2010}.
According to the plot of the relationship between $\tau_{(4,8)}[\psi(\theta)]$ and $\langle\psi(\theta)|B|\psi(\theta)\rangle$,
as the value of quantum violation increases the entanglement measure varies smoothly, ranging from the maximum 1 to the minimum $\sqrt{3}/2$.
In particular, we showed that there exist a family of four-qubit entangled states $\left|\psi(\theta)\right\rangle$ for which the WYKO inequality is drastically violated. Thus, the WYKO inequality can be acted as a strong entanglement witness for this family of four-qubit entangled state.

\section{Acknowledgements}
This work was supported by the National Natural Science Foundation of China under Grant Nos: 11475054, 11371005, 11547169, Hebei Natural Science Foundation of China under Grant No: A2014205060, the Research Project of Science and Technology in Higher Education of Hebei Province of China under Grant No: Z2015188, the Graduate Innovation Project of Hebei Province of China under Grant No: sj2015004, Langfang Key Technology Research and Development Program of China under Grant No: 2014011002.

\end{document}